\documentclass{llncs}
\usepackage{graphicx}
\usepackage{amsmath, amssymb, amsfonts}
\usepackage[latin1]{inputenc}
\usepackage[english]{babel}
\usepackage{enumerate}
\usepackage{ulem, cancel}
 
\usepackage{color}
\usepackage{subfigure}

\begin{document}

\title{Stochastic Hybrid Models of Gene Regulatory Networks -- A PDE Approach}

\author{Pavel Kurasov\inst{1}  \and Alexander L\"uck \inst{2}  \and Delio Mugnolo \inst{3} \and Verena Wolf \inst{2}}

\institute{Institute of Mathematics, Stockholm University, Stockholm, Sweden 
\and Department of Computer Science, Saarland University, Saarbr\"ucken, Germany
\and Department of Mathematics, FernUniversit\"at in Hagen, Hagen, Germany}

\maketitle

\begin{abstract}
 A widely used approach to describe the dynamics of gene regulatory networks  is based on the
chemical master equation, which considers probability distributions over all possible combinations of molecular counts. The analysis of such models is extremely challenging due to their large discrete state space.
We therefore propose a hybrid approximation approach based on a system of partial differential equations,
where we assume a continuous-deterministic evolution for the protein counts.
We discuss efficient analysis methods for both modeling approaches and 
  compare their performance. 
  We show that the hybrid approach yields accurate results for sufficiently large molecule counts, while reducing the computational effort from one ordinary differential equation for each  state to one partial differential equation 
  for each   mode of the system.
Furthermore, we give an analytical  steady-state solution of the    hybrid model for 
the case of a self-regulatory gene.
\\[1ex]
\end{abstract}

\section{Introduction}
\label{intro}

In the last decades biological measurements have become increasingly quantitative and 
have fostered new   approaches  for the analysis of  models that describe quantitative aspects of
biological systems. 
Moreover, it has been observed that cellular processes such as gene expression are
shaped by random events which led to an increasing interest in stochastic models 
(see, for instance, \cite{kaufmann2007stochastic}).
Therefore, in the last decade quantitative stochastic models have been widely
used to test and verify hypotheses about the   structure and function of 
biological systems on a microscopic scale.
In particular, chemical master equation (CME) models, that assume an underlying
discrete-state Markov process, are well-established for describing 
  gene regulatory networks \cite{Ovidiu}. 
As opposed to continuous-deterministic models, they use discrete variables to
count the number of molecules of each chemical species. 
In particular, such models take into account   low copy numbers, which are known to be the source
of  cellular stochasticity. 
Since CME models are  in most cases
too complex to be solved analytically, 
 approximative numerical analysis methods have been
developed \cite{didier2009fast,munsky2006finite,sidje2007inexact,wolf2010solving} as well as statistical approaches based on Monte-Carlo simulation \cite{gillespie1977exact}.
Exceptions are exact solutions for models that obey 
detailed balance \cite{Laurenzi2000} and for those that assume that all intracellular interactions 
are monomolecular \cite{jahnke}. 
Since gene regulatory networks typically contain feedback loops, second-order 
interactions are necessary to describe the  evolution of the system.
Moreover, neither   detailed balance   nor linear dynamics are realistic assumptions
  even for simple regulatory networks.
Recently,   analytical solutions for single-gene feedback loops have been presented~\cite{grima2012steady,hornos2005self,kumar2014exact,liu2016decomposition,vandecan2013self,visco2008exact}.

 Here, we propose a stochastic hybrid approach
 for gene regulatory networks, in which only the state of the genes is represented by
 a discrete-stochastic variable while we assume that for a fixed gene state, the evolution of the protein numbers is deterministic and described by an ordinary differential equation (ODE).
 Thus, we dismiss the detailed discrete-state description  of highly-abundant
 chemical substances in the cell and use a discrete-stochastic description only where it
 is really necessary, for instance, when boolean variables are used to describe whether a gene
 is active or not. More precisely, we assume that a gene  can stochastically switch between an 'on'   and an 'off' state and the switching probability depends on the global state of the system, i.e., it is a function of
 the (continuous) protein concentrations (or counts).
Hence, the models that we consider are a special case of
 piecewise deterministic Markov processes \cite{Davis1984} which have been successfully
 applied to gene regulatory networks in earlier work \cite{herbach2017inferring,Lin20170804,zeiser2008simulation}. 
  Our assumption about the continuous-deterministic protein dynamics  eases the derivation of exact solutions for the steady-state distribution of the process
  and is equivalent to the assumptions made in 
    stochastic  hybrid simulation algorithms for CME models \cite{crudu2009hybrid,herajy2012hybrid,marchetti2016hrssa,melykuti2014equilibrium,pahle2009biochemical,singh2005models,singh2010stochastic,zhang2008zero}.

Besides assuming   the two gene modes 'on' and 'off', we consider for each gene
a variable for the corresponding protein concentration. The rates at which the global mode
changes may depend on the global state  of the system, i.e., both the state of the genes and
the protein concentrations. 
Our model does not explicitly model transcription and the concentration of mRNA molecules. 
Instead, we assume that the  evolution of the concentration or count of a certain protein is determined by 
the  current (stochastic) state of the corresponding gene. 
More concretely,  we assume mode-dependent production rates for the proteins and
fixed degradation rates.
The evolution of the mode-conditional density functions describing the protein concentrations
 is then given by a system of 
(first-order) partial differential equations (PDE), which can be solved numerically.
A comparison to a numerical solution of the corresponding CME,
i.e. the equation for the same process except that   protein counts/concentrations are discrete
random variables  
following a Markov jump process description, shows that as long as the protein counts are not 
too small, the PDE gives a very accurate approximation of the underlying ``true'' probability distribution.
We present results for examples with slow and fast switching in one dimension as well as examples 
 with uni- and bimodal distributions in two dimensions to illustrate the  applicability of our approach.

Previous work on stochastic models of gene regulatory networks mostly focus on analytical solutions for fully discrete-stochastic descriptions \cite{grima2012steady,hornos2005self,visco2008exact}
 or on the hybrid 
sampling approaches mentioned above
\cite{crudu2009hybrid,herajy2012hybrid,marchetti2016hrssa,melykuti2014equilibrium,pahle2009biochemical,singh2005models,singh2010stochastic,zeiser2008simulation,zhang2008zero}.
For gene regulatory networks, that experience burst behavior in the protein production,  partial integral differential equations 
have successfully   been applied to describe bursting \cite{Friedman2006,PAJARO201751}. However, state changes
of genes are described by Hill functions.
 Here, we do not assume any burst behavior but concentrate on a generic discrete-stochastic description of the 
gene states in combination with  continuous-deterministic dynamics 
of the corresponding protein concentrations.
In addition, we consider the special case of a self-regulated gene, which represents a motif  that is often part of more complex networks. 
We present   a closed-form solution
of its steady-state density  and compare it to the   closed-form solution  proposed by~Grima et al. \cite{grima2012steady}
for the corresponding fully discrete CME model.
 The comparison shows that the distributions agree as long as the protein counts are large, i.e. at the order of hundreds.

Another class of hybrid models is based on a mixture of the CME approach and the linear-noise approximation.
There the CME is used to describe the behavior of the genes, while the linear-noise approximation describes the processes involving the proteins \cite{Hufton2016,Thomas6994}.

A recent review on hybrid and non-hybrid methods of stochastic simulation in biology can be found in \cite{Schnoerr2017}.

 The paper is organized as follows:
In Section~\ref{sec:2} we introduce the CME and our hybrid modeling approach for the description of gene regulatory networks.
In Section~\ref{sec:4} the analytical solution for a special case of the hybrid model is compared to an analytical solution based on the CME.
The numerical solution of this model together with a case study are presented in Section~\ref{sec:3}.
Finally in Section~\ref{sec:5} we conclude our results.

\section{Stochastic Models}
\label{sec:2}
In the following, we will  recapitulate the fully discrete CME approach which has widely been used for the description of gene regulatory networks. Then we will introduce the hybrid model in which protein concentrations
are described by continuous-deterministic variables.
\subsection{Chemical Master Equation}
Let $\nu$ be the number of genes in the network and assume that each gene is used to produce 
a different type of protein, such that there are in total $2\nu$ chemical species $S_1,\ldots ,S_{2\nu}$.
The state of the system at time $t\geq0$ is then given by the random vector $\vec{Y}(t)=(Y_1(t),\ldots,Y_{2\nu}(t))$, where  $Y_i(t)$ is a discrete random variable that describes the number of molecules of species $S_i$ at time $t$. 
The random vector $\vec{Y}(t)$ changes according to a set of $J$ chemical reactions  where
for $j\in\{1,2,\ldots,J\}$ reaction $R_j$ is described by a stoichiometric equation of the form
\begin{equation}
R_j~:~s^{(j)}_{1}S_{1}+\cdots+s^{(j)}_{2\nu}S_{2\nu}\stackrel{\kappa_j}{\longrightarrow}w^{(j)}_{1}S_{1}+\cdots+w^{(j)}_{2\nu}S_{2\nu}.
\label{Eq:react}
\end{equation}
The stoichiometric coefficients for reaction $R_j$ and species $i$ are denoted by $s^{(j)}_i\in\mathbb{N}$ for the reactants and  $w^{(j)}_i\in\mathbb{N}$ for the products of the reaction.
If a species is not involved in the reaction the corresponding stoichiometric coefficient is set to zero.
The stochastic rate constant for reaction $R_j$ is denoted by $\kappa_j$.\\
If the system is in state $\vec{y}$ at time $t$, the conditional probability that reaction $R_j$ occurs in the time intervall $[t,t+dt)$ is given by
\begin{equation}
\alpha_j(\vec{y})dt=P(R_j\text{~occurs~in~}[t,t+dt)|\vec{Y}(t)=\vec{y}).
\end{equation}
The corresponding propensity can be calculated as
\begin{equation}
\alpha_j(\vec{y})=\kappa_j\prod_{i=1}^{2\nu}\binom{y_i}{s^{(j)}_i},
\end{equation}
which is the product of the rate constant $\kappa_j$ and all possible combinations of reactants, which are required for the reaction.\\
Due to the chemical reaction $R_j$ the number of molecules for some species change such that we can define a state change vector $\vec{v}_j=\vec{w}^{(j)}-\vec{s}^{(j)}$, i.e. the state change vector if given by the difference of the molecular counts in the states before and after the reaction.
Note that each reaction determines a unique state change vector since the number of molecules involved is fixed and does not depend on the absolute molecule counts.\\
The probability of being in state $\vec{y}$ at time $t$, starting from an initital state $\vec{y}_0$ is denoted by $p(\vec{y},t)$. 
We are now able to describe the dynamics of the system via the CME
\begin{equation}
\frac{d}{dt} p(\vec{y},t) =\sum_{j: \vec{y}-\vec{v}_j^- \geq0}\alpha_j(\vec{y}-\vec{v}_j)p(\vec{y}-\vec{v}_j,t)-\alpha_j(\vec{y})p(\vec{y},t),
\end{equation}
where we sum over all possible reactions that either lead to (first term in sum) or can occur in state $\vec{y}$ (second term in sum).
Note that at any time only the current state determines the system's future evolution.
Therefore $(\vec{Y}(t))_{t\geq0}$ is a continuous-time Markov chain with $2\nu$-dimensional state space.

For  gene regulatory networks, we assume that for each gene $G_i$  and its corresponding protein $P_i$ 
we have the reactions
\begin{align*}
R^{(i)}_{1,2}~&:~G^0_i \mathrel{\mathop{\rightleftharpoons}^{\mathrm{\lambda}}_{\mathrm{\mu}}}G^1_i,&\qquad&
R^{(i)}_{4}~:~G^0_i\stackrel{a}{\rightarrow}G^0_i+P_i,\\
R^{(i)}_{3}~&:~P_i\stackrel{d}{\rightarrow}\emptyset,&\qquad&
R^{(i)}_{5}~:~G^1_i\stackrel{c}{\rightarrow}G^1_i+P_i.
\end{align*}
Reaction $R^{(i)}_{1}$ turns the gene 'on', while reaction $R^{(i)}_{2}$ turns the gene 'off'.
Here, we assume that $\lambda$ and $\mu$ may be functions of the protein counts of the same or an other gene.
The degradation of protein is shown in reaction $R^{(i)}_{3}$.
Reaction $R^{(i)}_{4}$ and $R^{(i)}_{5}$ correspond to protein production in different gene states, for example no/weak (state $G^0_i$) and strong (state $G^1_i$) production if $c>a$.
This yields a special case of the CME and for the case $\nu=1$, closed-form solutions 
for the steady-state solution of the CME have been derived \cite{grima2012steady}.
Moreover, software tools have been developed for the case that the molecular counts are not too large
and  a numerical integration of the CME is possible~\cite{kazeroonian2016cerena,lapin2011shave}.

\subsection{Stochastic Hybrid Model}
For the hybrid description of gene regulatory network with $\nu$ genes, we split the state
vector $\vec{y}$ into two coupled random vectors $\vec{m}=(m_1,\ldots,m_\nu)$
and $\vec{x}=(x_1,\ldots,x_\nu)$, where for   $i\in \{1,\ldots,\nu\}$ we define 
$m_i\in\{0,1\}$ as the state of gene $i$ and
$x_i$ as the corresponding concentration or number of proteins.
Here, $m_i=0$ represents the case where the gene is inactive and $m_i=1$ the case where it is active.
Since there are two possible states for each of the $m_i$, $i\in \{1,\ldots,\nu\}$, the total number of possible modes of the system is $M=2^{\nu}$.
Depending on the model not all   modes may be reachable
from some initial configuration. 
In the sequel, we will assign 
  enumeration index $z$ to mode $\vec{m}$ by converting the binary number $[m_{\nu}m_{\nu-1}\ldots m_2m_1]_2$  
  into a decimal number and adding one to ensure that $z\in \{1,\ldots,2^{\nu}\}$, i.e.   $z=[m_{\nu}m_{\nu-1}\ldots m_2m_1]_2+1$.

In our hybrid model, we assume that all protein concentrations change deterministically  according to some 
linear differential equation, whereas   transitions between   modes follow a Markov jump process.
The diagonal matrix $\boldsymbol{R}_i \in \mathbb{R}^{M\times M}$ describes the concentration change of protein $i$ in mode $z$ and has the form
\begin{align}
\boldsymbol{R}_i^{(z,z)}(\vec{x})=
\begin{cases}
a_i-b_i x_i,~&\text{if~}m_i=0, \\
c_i-d_i x_i,~&\text{if~}m_i=1,
\end{cases}
\label{Eq:R}
\end{align}
where $a_i$ and $c_i$ are the production rates of protein $i$ if the gene is inactive or active, respectively\footnote{It is possible to choose $a_i=0$ if no proteins are produced in the inactive state. Alternatively, a (weak)  production rate $a_i>0$ may be chosen. }.
The respective degradation rate constants are given by $b_i$ and $d_i$ and the 
corresponding degradation rates 
are proportional to the protein concentration. Although, 
the cases requiring $b_i\neq d_i$ may be rare, we do not restrict to the case $b_i=d_i$ here.  
The infinitesimal generator matrix $\boldsymbol{Q}  \in \mathbb{R}^{M\times M}$ describes the transitions between   different modes.
Consider, for example, an exclusive switch with two genes and a
 common promoter region~\cite{loinger2007stochastic}.
 At most one protein can  bind to the promoter region at a time
 and it represses the production of the other protein.
 Then the matrix has the form
\begin{equation}
\boldsymbol{Q}=
\begin{pmatrix}
0 & 0 & 0 & 0\\
0 & -\lambda_1 & 0 & \lambda_1\\
0 & 0 & -\lambda_2 & \lambda_2\\
~~~0~~~ & ~~~~~\mu_1~~~~~ & ~~~~~\mu_2~~~~~ & -(\mu_1+\mu_2)
\end{pmatrix},
\label{Eq:Q}
\end{equation}
where $\lambda_i$ and $\mu_i$ are the rates 
at which     gene $i$ switches from the inactive to the active
state and vice versa. 
However,  the first mode  where both genes are inactive is not reachable
from the other three modes.
Furthermore we also allow concentration dependent parameters, i.e. the entries
of $\boldsymbol{Q}$ are continuous functions in $\vec{x}$ and $t$. 

The model described above can be represented by a fluid stochastic Petri net (FSPN)  \cite{horton1998fluid,trivedi1993fspns} by considering the gene states as the discrete marking and the protein counts as the (continuous) fluid levels.
 The system's time evolution is given by the   linear first-order  hyperbolic    partial differential equation (PDE)
\begin{equation}
\frac{\partial}{\partial t}
\vec{f}(\vec{x},t)= -\sum_{i=1}^{\nu} \frac{\partial}{\partial x_i}\vec{f}(\vec{x},t)\boldsymbol{R}_i(\vec{x})+\vec{f}(\vec{x},t)\boldsymbol{Q}(\vec{x},t),
\label{Eq:origPDE}
\end{equation}
where $\vec{f}(\vec{x},t)=(f_1(\vec{x},t),\ldots,f_M(\vec{x},t))$ is the 
vector of mode probability densities.
Intuitively this equation can be derived from the conservation of probability mass with a corresponding balance equation.
The in- and outflow of probability mass from the continuous part (changes in protein counts) is encoded in the terms containing $\boldsymbol{R}$, while the term with $\boldsymbol{Q}$ describes the in- and outflow of probability mass from the discrete part (changes of gene states) of the model.
Note that by using Eq.~\eqref{Eq:origPDE} the complexity is reduced to one PDE per mode (in total $2^{\nu}$) of the underlying probability distribution instead of one ODE per state (up to $\tilde{N}^{2\nu}$, where $\tilde{N}$ is the maximum number of a protein species' count) when using a CME approach.

\section{Analytical Steady-State Solution for a Self-Regulated Gene}
\label{sec:4}
In this section, we present an analytical solution of the steady-state density
for
  the special case  of a single gene, i.e., $\nu=1$ and $M=2$.
Then,     Eq.~\eqref{Eq:origPDE} gives
\begin{align}
\begin{split}
\frac{\partial}{\partial t}
\begin{pmatrix}
f_1(x,t) \\ 
f_2(x,t)
\end{pmatrix}^\intercal
= -\frac{\partial}{\partial x}\left[
\begin{pmatrix}
f_1(x,t) \\ 
f_2(x,t)
\end{pmatrix}^\intercal
\begin{pmatrix}
a-bx & 0 \\ 
0 & c-dx
\end{pmatrix}
\right]\\
+
\begin{pmatrix}
f_1(x,t) \\ 
f_2(x,t)
\end{pmatrix}^\intercal
\begin{pmatrix}
-\lambda(x) & \lambda(x) \\ 
\mu(x) & -\mu(x)
\end{pmatrix}.
\label{Eq:PDE2D}
\end{split}
\end{align}
 We assume a general linear form for the binding and unbinding rate here, i.e. $\mu(x)=mx+n$ and $\lambda(x)=kx+l$. 
In the steady-state ($\partial_t \vec{f}=0$) Eq.~\eqref{Eq:PDE2D} becomes an ODE and can be rewritten in the form
\begin{equation}
\frac{\partial}{\partial x}
\begin{pmatrix}
\psi_1(x) \\ 
\psi_2(x)
\end{pmatrix}^\intercal
=
\begin{pmatrix}
\psi_1(x) \\ 
\psi_2(x)
\end{pmatrix}^\intercal
\begin{pmatrix}
-\lambda(x)+b & \lambda(x) \\ 
\mu(x) & -\mu(x)+d
\end{pmatrix}
\begin{pmatrix}
a-bx & 0 \\ 
0 & c-dx
\end{pmatrix}^{-1}\!\!\!\!\!,
\label{Eq:SteadyODE}
\end{equation}
where we replaced $\vec{f}(x,t)$ by $\vec{\psi}(x)$.
For an appropriate choice of the parameters and protein range, the righthandside of Eq.~\eqref{Eq:SteadyODE} is Lipschitz continuous and thus has a unique solution according to the Picard-Lindel\"of theorem.
A more detailed reasoning about the convergence to the steady-state solution for our system of hyperbolical PDEs and the derivation of the following analytical solution is provided in future work \cite{proofpaper}.  

The main idea of the derivation of the analytical solution is to show by adding the two components of Eq.~\eqref{Eq:SteadyODE} that for the steady-state solution $(\psi_1,\psi_2)$ it holds $(bx-a)\psi_1(x)=(c-dx)\psi_2(x)$ everywhere on $[\frac{a}{b},\frac{c}{d}]$.
By introducing the notation $h(x):=(bx-a)\psi_1(x)=(c-dx)\psi_2(x)$ and  subtracting the second component of Eq.~\eqref{Eq:SteadyODE} from the first one, one can deduce that $h$ satisfies an ordinary differential equation with Dirichlet boundary conditions, which can in turn be explicitly integrated.
This leads to the steady-state solution

\begin{align}
\begin{split}
\psi_1(x)&=\frac{K}{b}\exp\left(\left(\frac{l}{b}+\frac{m}{d}\right)x\right)\left(x-\frac{a}{b}\right)^{\left(\frac{al}{b^2}+\frac{k}{b}-1\right)}\left(\frac{c}{d}-x\right)^{\left(\frac{cm}{d^2}+\frac{n}{d}\right)},\\
\psi_2(x)&=\frac{K}{d}\exp\left(\left(\frac{l}{b}+\frac{m}{d}\right)x\right)\left(x-\frac{a}{b}\right)^{\left(\frac{al}{b^2}+\frac{k}{b}\right)}\left(\frac{c}{d}-x\right)^{\left(\frac{cm}{d^2}+\frac{n}{d}-1\right)},
\end{split}
\label{Eq:Analytic}
\end{align}
for $x\in[\frac{a}{b},\frac{c}{d}]$, with some constant $K\in(0,\infty)$. 
 By inserting Eq.~\eqref{Eq:Analytic} into Eq.~\eqref{Eq:SteadyODE} it is straightforward to show that $\psi_1$ and $\psi_2$ are indeed the steady-state solution of Eq.~\eqref{Eq:PDE2D}. 
The constant $K$ is chosen in such a way that
\begin{equation}
\int_{x_i}^{x_f} (\psi_1(x)+\psi_2(x))dx=1,
\end{equation}
where $x_i=\frac{a}{b}$ and $x_f=\frac{c}{d}$.
The marginal density of $x$ is then given by the sum of $\psi_1$ and $\psi_2$, i.e.
\begin{equation}
\psi(x)=\psi_1(x)+\psi_2(x).
\label{Eq:Psi}
\end{equation}
 Depending on the parameters, the functions in Eq.~\eqref{Eq:Analytic} show different limit behaviors at the boundaries (see also \cite{proofpaper}), namely  
\begin{enumerate}[(i)]
\item if $al + kb < b^2$, then $\psi_1$ is singular at $x_i=\frac{a}{b}$,
\item if $al + kb = b^2$, then $\psi_1$ attains a nonzero value at $x_i=\frac{a}{b}$,
\item if $al + kb > b^2$, then $\psi_1$ tends to zero at $x_i=\frac{a}{b}$,
\item if $cm + nd < d^2$, then $\psi_2$ is singular at $x_f=\frac{c}{d}$,
\item if $cm + nd = d^2$, then $\psi_2$ attains a nonzero value at $x_f=\frac{c}{d}$,
\item if $cm + nd > d^2$, then $\psi_2$ tends to zero at $x_f=\frac{c}{d}$.
\end{enumerate}
Assuming law of mass action kinetics leads to a linear binding and constant unbinding rate, i.e. $l=n=0$, such that conditions (i)-(iii) only compare the unbinding rate to the degradation rate in the 'off' state and conditions (iv)-(vi) the product of the production rate in the 'on' state and binding rate to the squared degradation rate in the 'on' state.
In general, it is biologically plausible that the degradation rates are much smaller than the rates for the other reactions, hence only to the conditions (iii) and (vi) are reasonable.
If $ x_i= 0$ and the unbinding is very slow too, then condition (iii) may be applicable instead of (ii), since negative protein counts are impossible.
Since biologically meaningful results should not contain singularities, conditions (i) and (iv) are obviously not applicable.

As a next step we compare the above results  with 
the analytic solution of the corresponding CME derived by Grima et al. \cite{grima2012steady}, where we used the Taylor series approach described in \cite{cao2018linear} to obtain the results of the CME in a fast and accurate way. 
Since we obtain a   density whereas the CME yields a discrete   distribution, we    discretize
 our solution to compute the Hellinger distance $H$ of the two distributions.
The discrete distribution $P(N)$ yields only probabilities for positive integer numbers $N$.
In order to include the information from the real $x$ values from the continuous solution, we calculate mean values around the integers as follows
\begin{equation}
\tilde{\psi}(N)=\int_{N-0.5}^{N+0.5}\psi(x)dx.
\label{Eq:Psi_tilde}
\end{equation}
The Hellinger distance is then given by
\begin{equation}
H=\frac{1}{\sqrt{2}}\sqrt{\sum_{n=1}^{\hat{N}}\left(\sqrt{\tilde{\psi}(n)}-\sqrt{P(n)}\right)^2}
\label{Eq.Hellinger}
\end{equation}
where we use the notation $\hat{N}=\lfloor\frac{c}{d}\rfloor$.

To compare our solution to that of Grima et al., we match
the parameters in \cite{grima2012steady} as follows:
\begin{align*}
r_b& =a&
r_u& =c&
s_b& =m&
s_u& =k&
k_f& =b=d
\end{align*}
Without loss of generality, we consider only  the case $c>a$.
 Furthermore   our model does not allow   degradation of bound proteins, hence $k_b$ from \cite{grima2012steady} is set to $0$. Note that since at most a single protein can be bound and we focus 
 on systems with moderate or high protein numbers, this assumption
 is reasonable. 
Moreover, as in  \cite{grima2012steady} we assume linear binding rates 
and constant unbinding rates and therefore set $l=n=0$. 
In the $a>c$ case, $a$ and $c$, $m$ and $l$ as well as $k$ and $n$ swap roles, but the analytical solution~\eqref{Eq:Analytic} no longer holds. 

\begin{figure}[tb]
\centering
\subfigure[]{\includegraphics[scale=0.29]{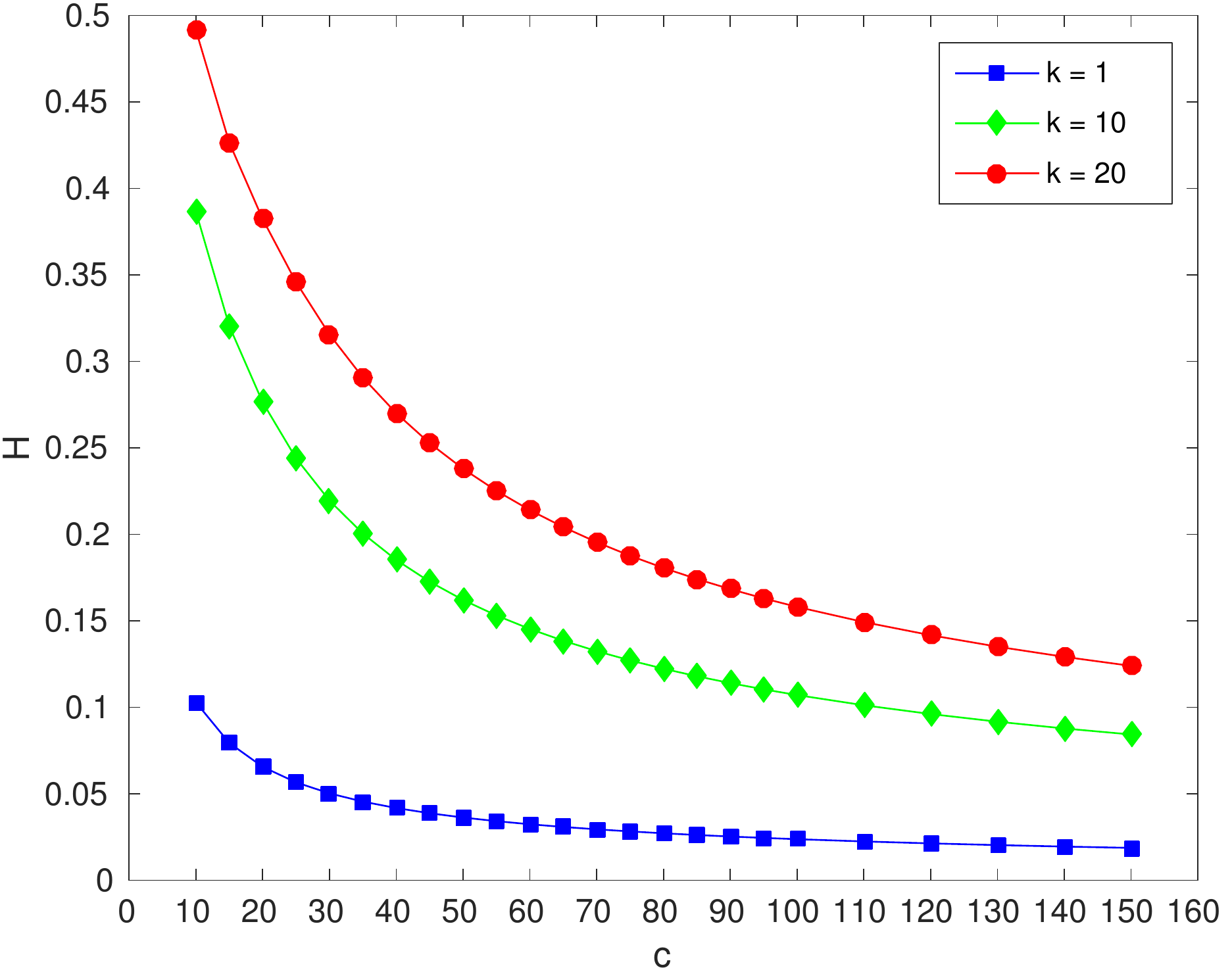}}\hspace{0.5cm}
\subfigure[$c\!=\!50,k\!=\!1,m\!=\!1$]{\includegraphics[scale=0.42]{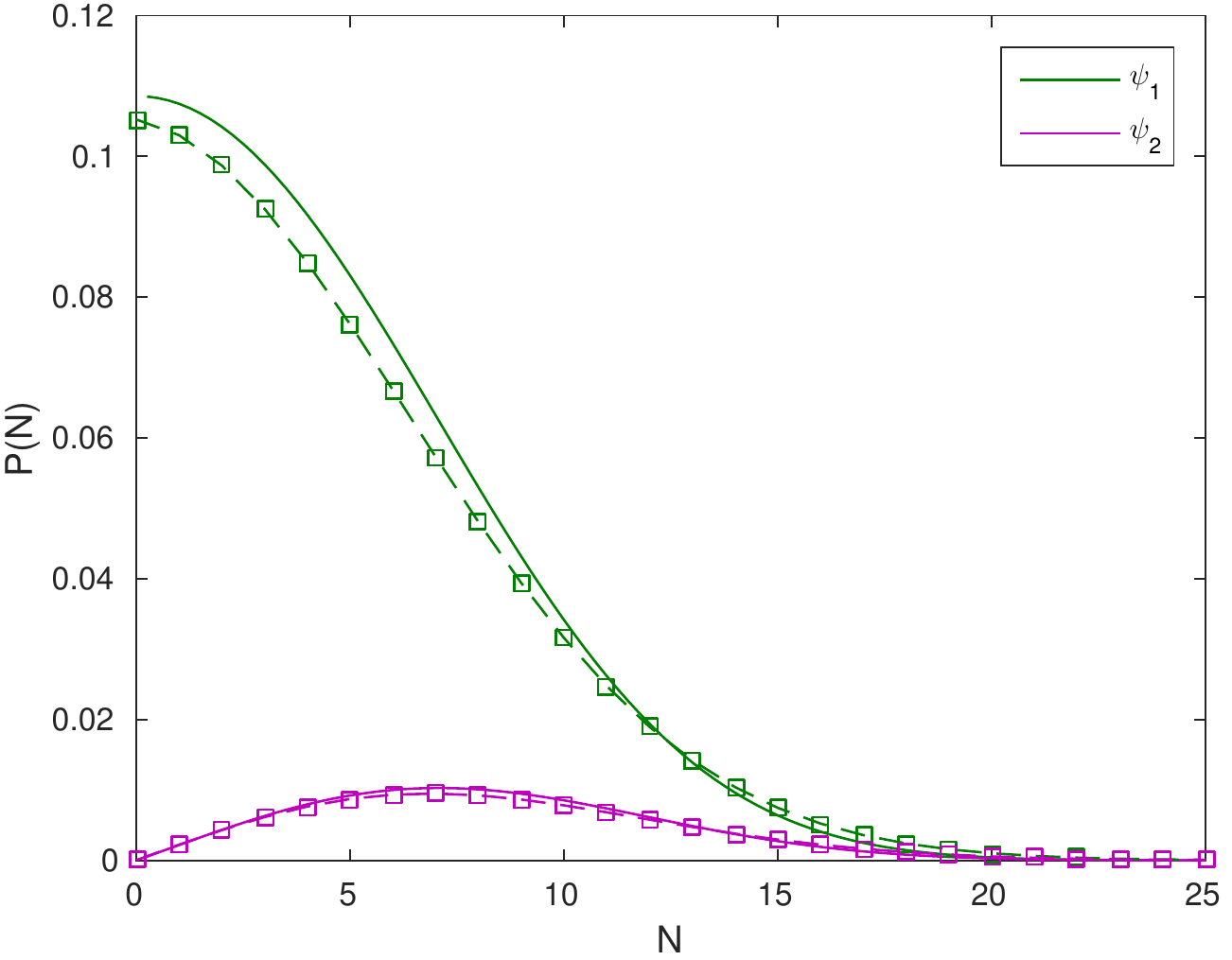}}\\
\subfigure[$c\!=\!50,k\!=\!20,m\!=\!1$]{\includegraphics[scale=0.42]{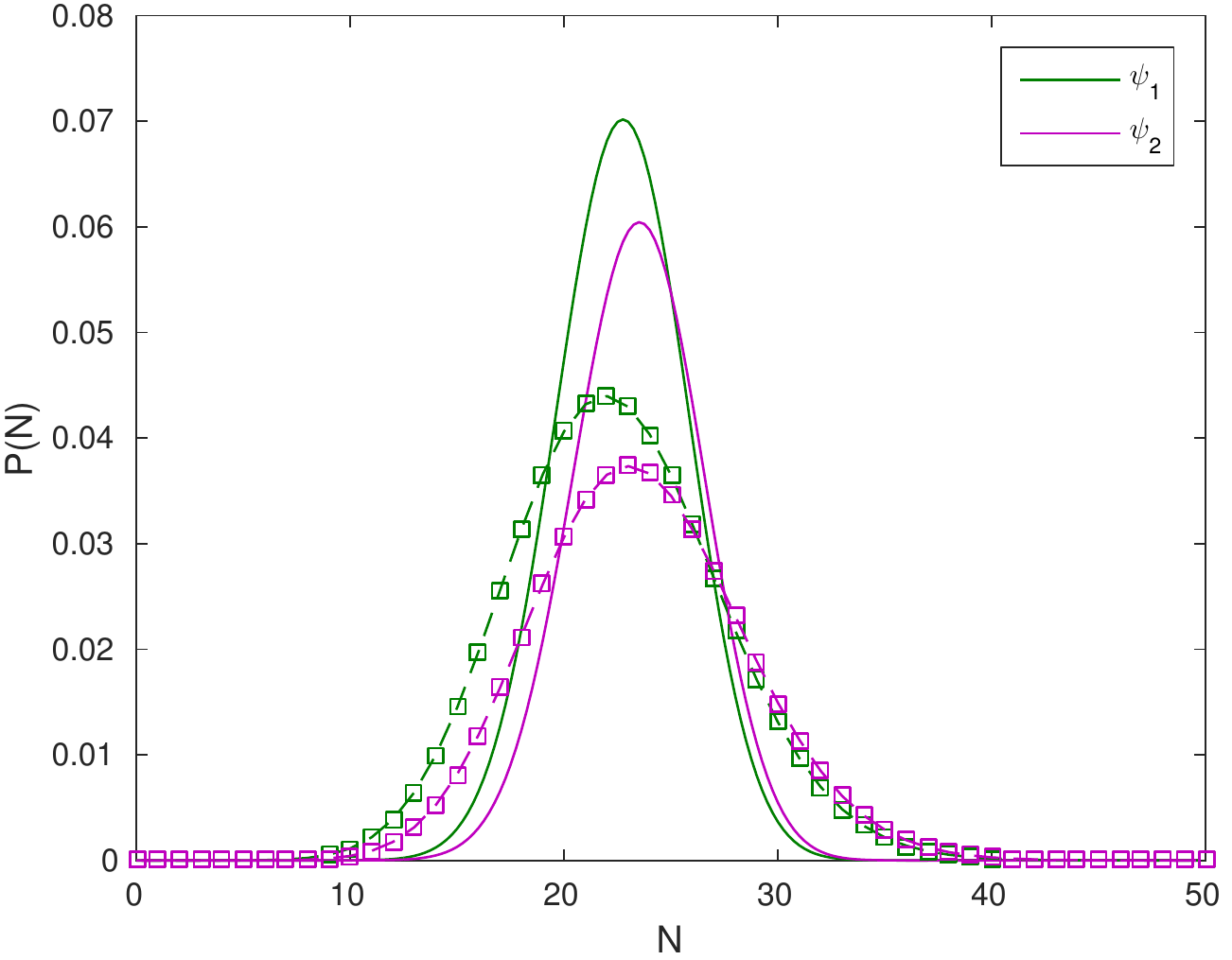}}\hspace{0.5cm}
\subfigure[$c\!=\!5000,k\!=\!200,m\!=\!0.1$]{\includegraphics[scale=0.42]{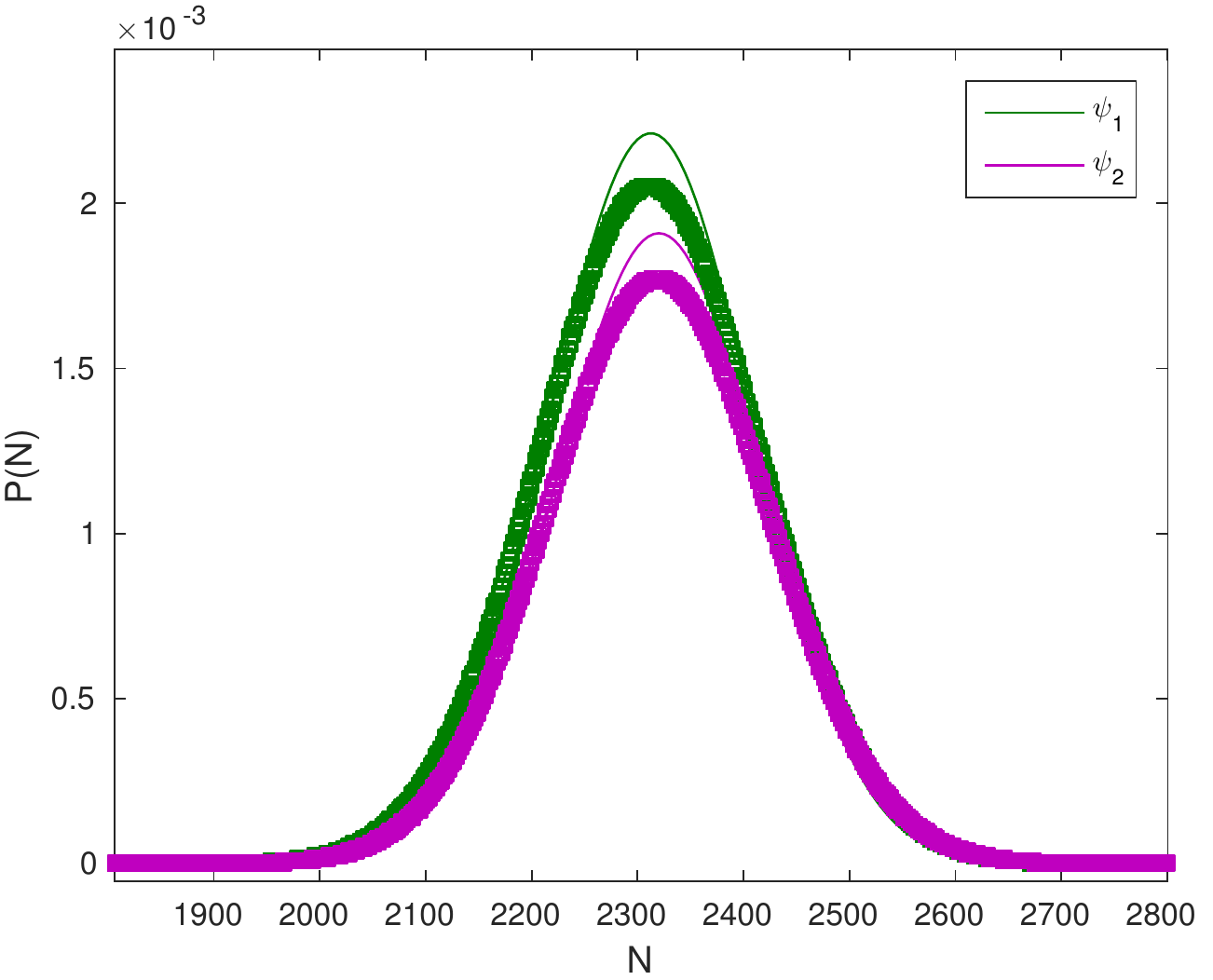}}\\
\caption{(a) Hellinger distance $H$ between \eqref{Eq:Psi_tilde} and the solution of \cite{grima2012steady}
for varying protein production parameter $c$.
(b), (c) Probability distribution and density for \eqref{Eq:Analytic} (solid lines) and \cite{grima2012steady} (markers) with varying parameters $c,k,m$.   (d) Similar to (b), (c) but the discrete distribution 
was computed using the tool SHAVE \cite{lapin2011shave}. 
\label{Fig:1DComp}}
\end{figure}

The plot in Fig.~\ref{Fig:1DComp}~(a)  shows the Hellinger distance $H$ between the two models for three different parameter sets in dependence of $c$ for three different unbinding rates.
The remaining parameters are $a=0$, $m=1$ and $d=1$. For increasing $c$ the average number of particles also increases, while $H$ decreases.
That means that with a larger protein number the distributions become more and more similar.  

 Fig.~\ref{Fig:1DComp}~(b)-(d) shows the probability distributions
 and densities for different choices of the parameters $c,k,m$, where the solid lines correspond to the density
 in Eq. \eqref{Eq:Analytic} and those with markers to the discrete
 probability distribution of the CME. For the  case of
 small protein counts in (b) and (c) we evaluated the solution 
 given in \cite{grima2012steady}. For the case of large protein 
 counts in (d) we used the tool SHAVE
to solve the CME~\cite{lapin2011shave} numerically until
steady-state. 
The distribution for the first mode is shown in dark green, the distribution for the second mode in purple.
 The remaining parameters are fixed as $a=0$ (no production
 if the gene is inactive) and   $d=1$ (degradation of proteins)
 for all examples.
 We see in Fig.~\ref{Fig:1DComp}~(b) that the solutions are also more similar for slower switching rates even for small protein counts.
  This is because switching between the two modes seems to
   increase the influence of the 
  fluctuations of the protein counts on the  joint discrete
  probability
  distribution that results from the solution of the CME.  
Note that the results of SHAVE and those based on \cite{grima2012steady} are nearly identical for small protein numbers. For large numbers, as in (d), only the SHAVE solution is shown. 
Also note that the variance of the density is lower compared to that of the discrete distribution. This holds also for transient solutions as shown in the next section and comes from the 
fact that we assume deterministic continuous dynamics for the
protein counts.

\section{Numerical Solution}
\label{sec:3}
Since $\boldsymbol{R}_i(\vec{x})$  as defined in Eq.~\eqref{Eq:R}  is a diagonal matrix,  Eq.~\eqref{Eq:origPDE}   for the mode with index $z$ can be written in the form of a transport equation
\begin{equation}
\frac{\partial}{\partial t}f_z(\vec{x},t)+\sum_{i=1}^{\nu} v_i(\vec{x})\frac{\partial}{\partial x_i}f_z(\vec{x},t)=g_z(\vec{x},t),
\label{Eq:TransportPDE}
\end{equation}
where $v_i(\vec{x})=R_i^{(z,z)}(\vec{x})$ is the transport velocity in $i$ direction and 
\begin{equation}
g_z(\vec{x},t)=f_z(\vec{x},t) \cdot \left(-\sum_i \frac{\partial}{\partial x_i}R_i^{(z,z)}(\vec{x})\right)+\vec{f}(\vec{x},t){\boldsymbol{q}_z}(\vec{x},t)
\end{equation}
 contains all the remaining terms which do not include derivatives. Here, ${\boldsymbol{q}_z}$ is the $z$-th column of $\boldsymbol{Q}$.

To solve the above transport equation numerically, we employ a \emph{finite differen\-ces} scheme and divide the time interval $[0,T]$ and the intervals of protein counts $[x_i^{\text{min}}, x_i^{\text{max}}]$ into subintervals of
fixed length $dt$ and $\Delta$, respectively, where $T$ is the total simulation time and $x_i^{\text{min}}$ and $x_i^{\text{max}}$ form a sufficiently large range in the protein count for protein $i$. 
Obviously, it is also possible to apply more sophisticated discretization schemes based on variable interval lengths. However, for the examples that we considered equally spaced intervals yielded sufficiently accurate and fast results. 
For each subinterval we assume that the respective variable has a constant value 
 equal to the left interval boundary.
To simplify the notation, we consider only the case with   one protein ($\nu=1$)  
and write $x_\ell$ for the protein concentration that corresponds to the $\ell$-th interval.
We use the forward difference to approximate the time derivatives for the $r$-th interval $[t_r, t_{r+1}]$.
 To approximate the spatial derivatives we use a so-called \emph{Upwind scheme}, i.e. depending on the sign of $v_i(\vec{x})$ either forward or backward differences are used in order to take the different transport directions into account and obtain a numerically stable solution \cite{courant1952solution}.
Hence the spatial derivatives are approximated by 
\begin{align}
\begin{split}
\frac{\partial}{\partial x} f_z(x,t_r)|_{x=x_\ell} \approx&\\ 
\frac{\Delta f_z(x_{\ell},t_r)}{\Delta}=&
\begin{cases}
 &\dfrac{f_z(x_{\ell},t_r)-f_z(x_{\ell-1},t_r)}{\Delta},~\text{if~}v(x_{\ell})>0,\\[0.25cm]
 &\dfrac{f_z(x_{\ell+1},t_r)-f_z(x_{\ell},t_r)}{\Delta},~\text{if~}v(x_{\ell})<0.
\end{cases}
\end{split}
 \label{Eq:derSpace}
\end{align}
 Given some initial and boundary conditions and inserting the approximations for the derivatives into Eq.~\eqref{Eq:origPDE} the PDE can then be solved by the recursion scheme
\begin{align}
\begin{split}
\vec{f}(x_\ell,t_{r+1})=\vec{f}&(x_\ell,t_r)\\&-dt\left(\frac{\Delta \vec{f}(x_{\ell},t_r)}{\Delta}\vec{R}(x_\ell)+\vec{f}(x_\ell,t_r)\left[\vec{R}'(x_\ell)-\vec{Q}(x_{\ell},t_r)\right]\right)
\end{split}
\label{Eq:NumScheme}
\end{align}
Note that since $\vec{R}(x)$ is known, the derivative $\vec{R}'(x)$ can directly be calculated and  an approximation via finite differences is not needed here.
Moreover, the generalization of Eq.~\eqref{Eq:NumScheme} for multiple spatial variables is straightforward. 
Also, if the interval for the protein count is chosen large enough, suitable boundary conditions for the numerical solution  are $f_z(x_b,t)=0$ for all $t$, with $x_b=x^{\text{min}}$ and $x_b=x^{\text{max}}$.
Since $\vec{f}(x,t)$ is a probability density, 
we always find some $x^{\text{min}}$ and $x^{\text{max}}$, such that $f_z(x,t)<\varepsilon$ for $x<x^{\text{min}}$ or $x>x^{\text{max}}$.
The same considerations remain true for more than one protein species, i.e. for more than one spatial variable.

\subsection*{Case Study: Exclusive Switch}
\begin{figure}[tb]
\centering
\includegraphics[scale=0.4]{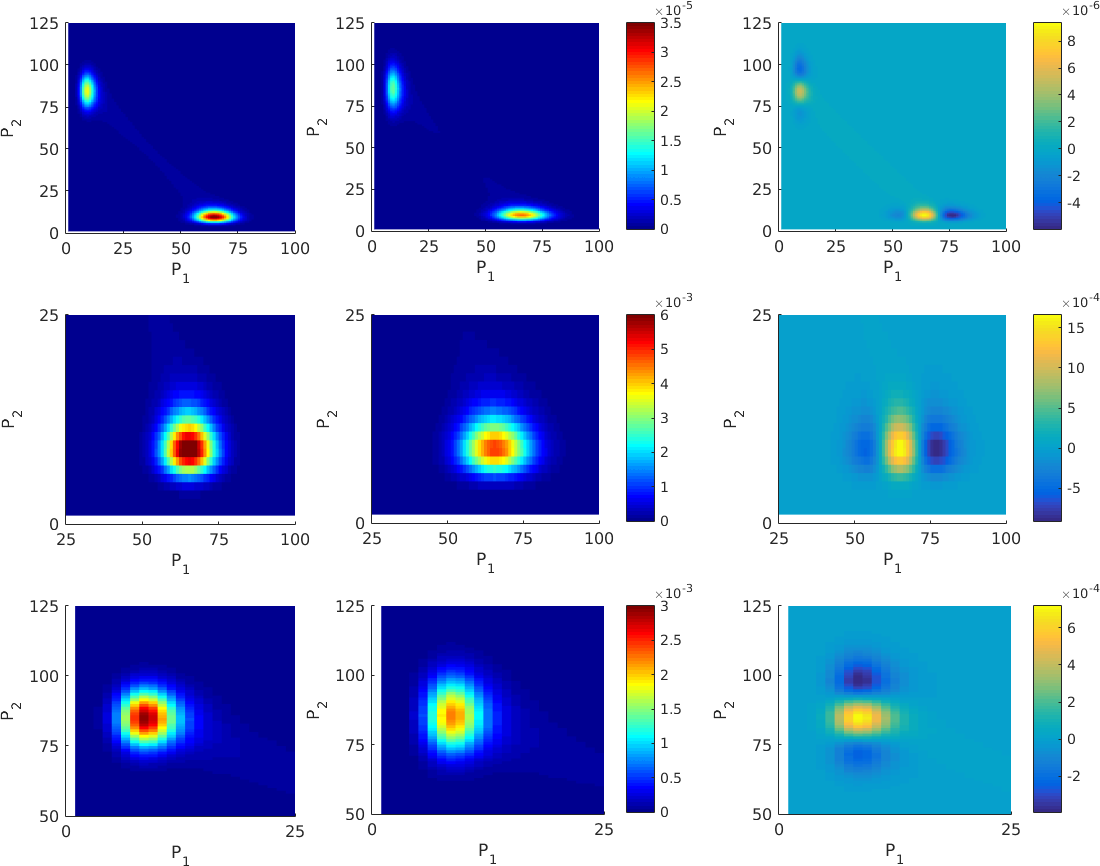}
\caption{Left: Numerical solution of the PDE system \eqref{Eq:PDEsys} in the bimodal case. Middle: Discrete probability distribution of the CME. Right: Difference of the solutions. Each row shows the results for one of the three reachable modes, i.e. promoter free (row 1), protein 1 bound (row 2) and protein 2 bound (row 3). The corresponding parameters are listed in Tab.~\ref{Tab:Paras2D}.  Note that different ranges for $P_1$ and $P_2$ are shown for the different modes. \label{Fig:2Dbimod}}
\end{figure}

\begin{figure}[tb]
\centering
\includegraphics[scale=0.4]{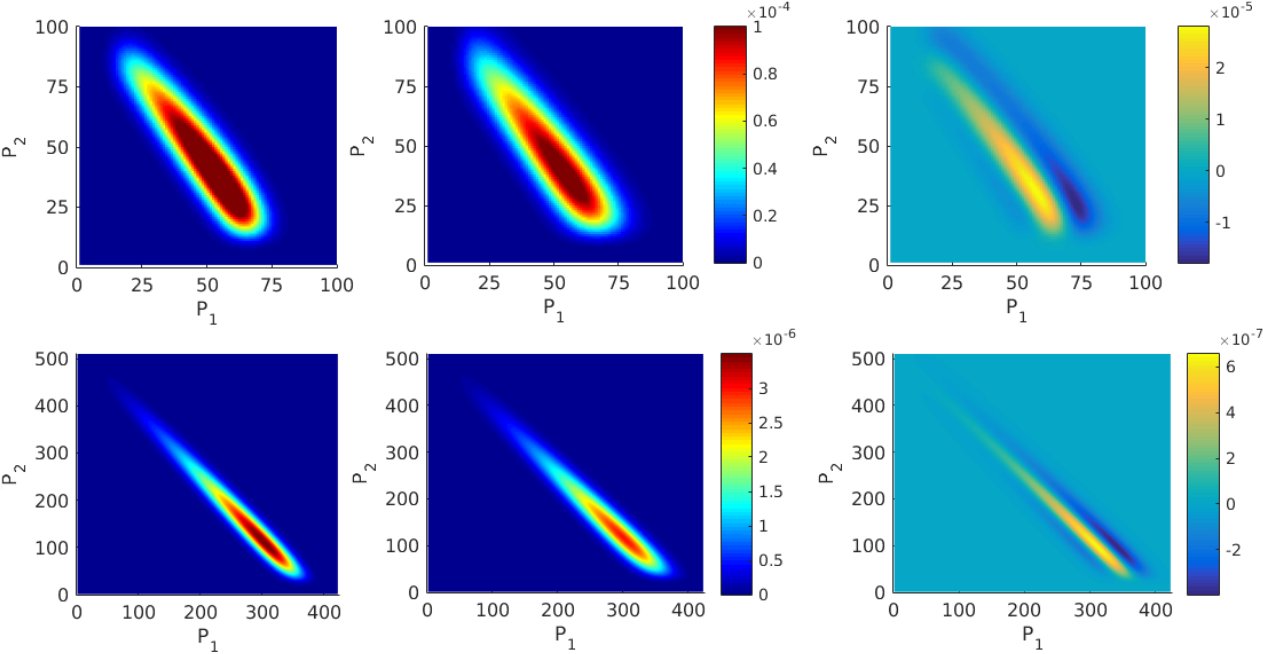}
\caption{Left: Numerical solution of the PDE system \eqref{Eq:PDEsys} in the unimodal case. Middle: Discrete probability distribution of the CME. Right: Difference of the solutions. Each row shows the promoter free mode for different average protein numbers. In the first row the average protein number is about $100$ proteins whereas the average protein number in the second row is about $500$ proteins in total. The corresponding parameters are listed in Tab.~\ref{Tab:Paras2D}.\label{Fig:2Dunimod}}
\end{figure}

We consider the exclusive 
  switch model above (see Eq.~\eqref{Eq:Q}) to compare the numerical solution of the hybrid model
  given by  Eq.~\eqref{Eq:origPDE}   
  to that of the corresponding CME model.
  For the former, we assume that if gene $i$ is active,  proteins of type $i$ are    produced at rate $c_i$.
Independent of the mode, the proteins of gene $i$ degrade at rate $d_i$, which we assume to be equal for all
three modes.
In the following we omit the   first  mode, which is not reachable, and set $M=3$, i.e. we remove the 
corresponding row and column  of zeros in $\bold{Q}$.
Then $f_2$ ($f_3$) corresponds to the mode where a protein of type 1 (type 2) is bound to the promoter, respectively, and  $f_4$ to mode where the promoter is free.
We assume no protein production if a gene is  not active, i.e., $a_1=a_2=0$ 
Furthermore,  we assume that the binding rate $\mu_i$ is proportional to the corresponding number of proteins, 
i.e., $\mu_i(\vec{x})=m_i x_i$. 
On the other hand, the rate $\lambda_i=k_i$ at which a protein of type $i$ unbinds from the promoter is independent of $\vec{x}$. 
With these assumptions the system of PDEs in Eq.~\eqref{Eq:origPDE} has the form
\begin{align}
\begin{split}
\frac{\partial}{\partial t}f_2(\vec{x},t)=&-\left((c_1-d_1x_1)\frac{\partial}{\partial x_1}f_2(\vec{x},t)-d_2x_2\frac{\partial}{\partial x_2}f_2(\vec{x},t)\right)\\
&+(d_1+d_2-k_1)f_2(\vec{x},t)+m_1 x_1 f_4(\vec{x},t),
\end{split}\nonumber\\
\begin{split}
\frac{\partial}{\partial t}f_3(\vec{x},t)=&-\left(-d_1x_1\frac{\partial}{\partial x_1}f_3(\vec{x},t)+(c_2-d_2x_2)\frac{\partial}{\partial x_2}f_3(\vec{x},t)\right)\\
&+(d_1+d_2-k_2)f_3(\vec{x},t)+m_2 x_2 f_4(\vec{x},t),\label{Eq:PDEsys}
\end{split}\\
\begin{split}
\frac{\partial}{\partial t}f_4(\vec{x},t)=&-\left((c_1-d_1x_1)\frac{\partial}{\partial x_1}f_4(\vec{x},t)+(c_2-d_2x_2)\frac{\partial}{\partial x_2}f_4(\vec{x},t)\right)\\
&+(d_1+d_2-(m_1 x_1+m_2 x_2))f_4(\vec{x},t)+k_1 f_2(\vec{x},t)+ k_2f_3(\vec{x},t).
\end{split}\nonumber
\end{align}
We consider  three   sets of parameters listed in Tab.~\ref{Tab:Paras2D} yielding  one bimodal and two unimodal densities.
Parameters that do not occur in the table, namely $a_i$, $l_i$, $n_i$, are set to $0$.
\begin{table}[bt]
\caption{Parameters of the exclusive switch used for the comparison in Figs.~\ref{Fig:2Dbimod} and \ref{Fig:2Dunimod}. \label{Tab:Paras2D}}
\centering
\renewcommand{\arraystretch}{1.25}
\begin{tabular}{|c|c|c|c|c|c|c|c|}
\hline 
~ & $c_1$ & $~c_2~$ & $d$ & $m_1$ & $m_2$ & $k_1$ & $k_2$ \\
\hline
bimodal & ~~~0.75~~~ & ~~~1.0~~~ & ~0.005~  & ~~0.02~~ & ~~0.01~~ & ~0.008~ & ~0.008~ \\
\hline
unimodal I& 0.75 & 1.0 & 0.005 & 0.01 & 0.01 & 0.1 & 0.2 \\
\hline 
unimodal II& 4.5 & 6.0 & 0.005 & 0.06 & 0.06 & 0.6 & 1.2 \\
\hline 
\end{tabular}
\end{table}
The results of the numerical PDE solutions using  the recursion scheme from Eq.~\eqref{Eq:NumScheme} generalized to two spatial variables   are plotted in the left column of Figs.~\ref{Fig:2Dbimod} and \ref{Fig:2Dunimod}. We choose $dt=10^{-2}$ for the approximation of the time derivative and $\Delta=1$ for both spatial derivatives. 
The plots in the column in the middle of Figs. \ref{Fig:2Dbimod} and \ref{Fig:2Dunimod} show the 
distribution obtained from a purely discrete model, i.e. when we solve the corresponding CME numerically. 
 We used the tool SHAVE  for the numerical integration of the CME, which 
 is based on a dynamical truncation of the state space \cite{lapin2011shave}.
 Note that it is also possible to obtain the distribution of the discrete model by 
 generating a large number of trajectories via Gillespie simulation.
 
We used an initial protein count of $10$ proteins per species and numerically simulated until $t=100$. 
 The right column shows the absolute difference   between the  numerical solution of the PDE and the CME distribution     at each point.

Note that the solution of the hybrid model shows a lower variance compared to the solution of 
the CME, which is reasonable
since in the PDE model proteins are assumed to change continuously and deterministically over time.
The randomness introduced by the (randomly occurring)  protein production and degradation events 
in the CME model is not taken into account in the hybrid model.

 We also applied our PDE approach to larger models, i.e. models with more species. An example for a gene regulatory network with three species is the so-called repressilator \cite{elowitz2000synthetic}. 
 For such networks, we got very similar results (not shown),
 i.e., the computed PDE densities gave accurate approximation of the CME distributions in case of moderate to high protein counts or slow switching rates.

Note that an extension to larger grid sizes ($\Delta>1$) of the above simple numerical solution scheme is straightforward, i.e. each cell represents a certain protein number range and, where the protein number occurs explicitly in the equations, an average protein number is considered. 
 For large protein numbers this aggregation is meaningful since neighboring states show a very similar behavior.
Note that 
 for $\Delta=1$ we have to integrate
the same number of equations  as the CME,  
while for $\Delta>1$ we have to integrate significantly less equations
in our hybrid approach.
Hence for $\Delta=2$ and $\Delta=3$ the running time decreases while the accuracy remains high (see Tab.~\ref{Tab:Runtime}).
We expect that with an adaptive grid, in which the cell size depends on the 
amount of probability mass and in which the grid changes over time, larger speed-ups 
at high accuracy are possible. 
The underlying idea to    reduce the number of equations in regions containing low probability mass
is also used by SHAVE (see [20]).
For our PDE solution, however, this requires
a more sophisticated implementation, in which merging and splitting 
of cells over time is possible, and is omitted here.

\begin{table}[tb]
\caption{Performance comparison of a direct numerical solution of the CME using SHAVE and the PDE solution with $dt=10^{-2}$ for different grid sizes $\Delta$ for the exclusive switch model with parameter set ``unimodal~II'' from Tab.~1. The Hellinger distance $H$ is calculated with respect to 
 the CME solution provided by SHAVE.\label{Tab:Runtime}}
\centering
\renewcommand{\arraystretch}{1.25}
\begin{tabular}{|c|c|c|c|c|}
\hline 
~ & ~~SHAVE~~ & ~~~$\Delta=1$~~~ & ~~~$\Delta=2$~~~ & ~~~$\Delta=3$~~~\\
\hline
~~runtime~~ & $407$~s & $1090$~s & $240$~s & $102$~s \\
\hline
$H$& $0$ & $0{.}0936$ & $0{.}1023$ & $0{.}2245$\\
\hline  
\end{tabular}
\end{table}

\section{Conclusion}
\label{sec:5}
We proposed a modeling framework for gene regulatory networks
that assumes  discrete random changes of the different gene states  and
continuous-deterministic changes for the protein concentrations.
For moderate and large protein counts or models with slow mode switching, the corresponding 
PDE solution yields accurate results. 
For the steady-state solution, we have a closed-form solution
which can efficiently be evaluated while previous approaches
suffer from numerical problems in the case of large protein counts.  
 We also presented a numerical scheme to compute transient solutions
 of the PDE. 
 
 For future work, we plan to investigate different
 numerical methods, in particular, adaptive discretization schemes. In addition, we will work on closed-form expressions for the steady-state solution for more complex networks, e.g. with two 
 genes and their corresponding proteins.   

\newpage
\subsection*{Acknowledgements}
The work of D.M. was partially supported by the Deutsche Forschungsgemeinschaft (Grant 397230547).  All four authors were partially supported by the Center for Interdisciplinary Research (ZiF) in Bielefeld, Germany, within the framework of the cooperation group on ``Discrete and continuous models in the theory of networks''.


\end{document}